\documentclass{jaa}

\usepackage{graphicx}
\usepackage[dvipsnames]{color}

\begin{document}

\title{Classical orbital paramagnetism in non-equilibrium steady state}

\author{Avinash A. Deshpande\textsuperscript{1} \and 
        N. Kumar\textsuperscript{1}}
\affilOne{\textsuperscript{1}Raman Research Institute, Bangalore 560080, India.}


\twocolumn[{

\maketitle

\corres{desh@rri.res.in}

\msinfo{25 March 2017}{20 July 2017}{31 July 2017}

\begin{abstract}
{We report the results of our numerical simulation of  
classical-dissipative dynamics of a charged particle subjected 
to a non-markovian stochastic forcing. We find that the system 
develops a steady-state orbital magnetic moment in the  
presence of a static magnetic field. Very 
significantly, the sign of the orbital magnetic moment 
turns out to be {\it paramagnetic} for our choice of parameters, 
varied over a wide range. This is shown specifically for the 
case of classical dynamics driven by a 
Kubo-Anderson type non-markovian noise. Natural spatial boundary 
condition was imposed through (1) a soft (harmonic) 
confining potential, and (2) a hard potential, approximating 
a reflecting wall. There was no noticeable qualitative difference. 
What appears to be crucial to the orbital magnetic effect noticed 
here is the non-markovian property of the driving noise chosen.  
Experimental realization of this effect on the laboratory scale, 
and its possible implications are 
briefly discussed. We would like to emphasize that the above steady-state 
classical orbital paramagnetic moment complements,  
rather than contradicts the Bohr-van Leeuwen (BvL) theorem 
on the absence of classical orbital diamagnetism in thermodynamic equilibrium.}
\end{abstract} 
 

}]


\doinum{12.3456/s78910-011-012-3}
\artcitid{\#\#\#\#}
\volnum{123}
\year{2017}
\pgrange{20--24}
\setcounter{page}{20}
\lp{24}


\section{Introduction}
The Bohr-van Leeuwen (BvL) theorem on the absence of classical orbital 
diamagnetism in thermodynamic equilibrium has been a surprise of 
theoretical physics 
(Bohr 1911; van Leeuwen 1921; van Vleck 1932; Peierls 1979; Ma 1985).
This BvL null result is statistical-mechanically exact. 
It is, however, counter-intuitive inasmuch as a charged particle orbiting 
classically under the Lorentz force exerted by an externally applied magnetic 
field is equivalent to an amperean current loop in the interior of the sample, 
and hence to a non-zero orbital magnetic moment. Moreover, the orbital moment 
is expected to be diamagnetic following the Lenz's law. A  physically appealing 
resolution was advanced by Bohr (1911)
suggesting an exact cancellation, 
on the average,  of the diamagnetic orbital magnetic moment in the interior 
of the sample by the paramagnetic moment subtended by the particle skipping 
the boundary in the opposite (paramagnetic) sense $-$ the edge current 
(Ma 1985; van Vleck 1932).

The BvL theorem on the absence of classical orbital (dia-)magnetism 
was re-examined  by Kumar \& Kumar (2009), where they 
considered dynamics of the charged particle in a finite but unbounded space,
namely the surface of a sphere (recall that, strictly speaking, a boundary
has no boundary).
They had found, through numerical simulations, 
that an equilibrium non-zero orbital diamagnetism 
could possibly  exist for the case of a charged-particle  motion
on the surface of a sphere, but not a plane.

Soon thereafter, in Pradhan \& Seifert (2010),
the role of the boundary was treated explicitly through 
a solution of the Fokker-Planck equation associated with the classical 
Langevin dynamics of the charged particle 
on the surface of a sphere.
Again, the orbital magnetic moment turned out to be zero. 

\begin{figure} [!t]
\centerline{\includegraphics[width=1.5\columnwidth]{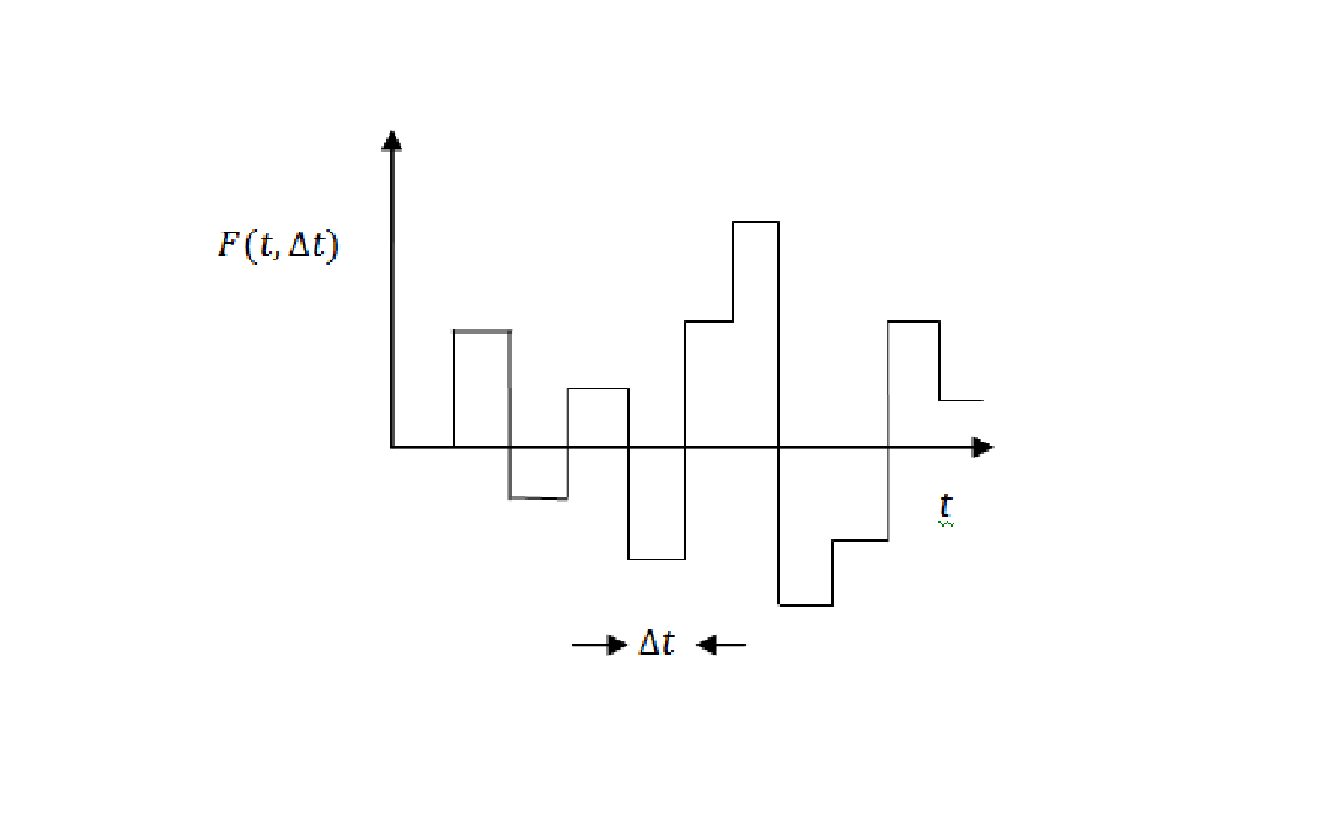}}
\caption{Shows schematically one possible realization of such 
a non-markovian noise comprising equi-spaced rectangular pulses 
of random gaussian height modeled on a Kubo-Anderson process.}
\end{figure}

More recently,  however, Deshpande, Kumar \& Kumar (2012) 
negated the result  of Kumar \& Kumar (2009) through similar numerical 
simulation, but now with finer time steps, for the motion of 
the charged particle on the surface of a sphere --in fact, 
in the long-time limit (i.e. in thermal equilibrium) 
the orbital moment indeed again turned out to vanish with 
decreasing time step, just as per the BvL Theorem! This trend 
revealed the possible role of finite temporal 
correlation scale of random forcing in inducing non-zero 
steady-state orbital magnetic moment.

Our analyses strongly suggest that the vanishing of the classical orbital 
diamagnetism is a direct consequence of detailed-balance (the microscopic 
reversibility), namely that there are {\it no cycles} in thermodynamic 
equilibrium. Now, the detailed balance is, of course, conditioned 
mathematically by the second fluctuation-dissipation (II-FD) theorem of 
Kubo (Kubo 1954,1966; Anderson 1954).
In terms of the classical Langevin equation governing the  
stochastic dissipative dynamics, that underlies equilibrium 
statistical mechanics, the F-D relation requires the driving noise 
to be markovian. Indeed, it has been shown (Kumar 2012)
that a markovian steady 
state can always be transformed to an equivalent thermodynamic state 
in equilibrium. It is our conjecture, therefore, that a stochastic 
dissipative system driven by a non-markovian noise may have in general 
a non-equilibrium steady state with finite orbital magnetic moment $-$ 
without conflicting with the BvL theorem for thermodynamic equilibrium. 
In what follows, we have addressed this issue through a detailed numerical 
simulation of the relevant stochastic dissipative dynamics involved. 
The results of our simulation support our conjecture, namely that the 
orbital magnetic moment is indeed finite for the non-markovian case. 
There is, however, a new surprise now, namely that the sign of the 
magnetic moment turns out to be paramagnetic! Moreover, for a 
classical gas of such charged particles with paramagnetic orbital 
moment, the inherently positive feedback may lead to an enhanced 
magnetic susceptibility $-$ possibly even to a spontaneous ordering 
of the classical orbital magnetic moments. 

\section{Stochastic dissipative dynamics in a magnetic field}
Consider the stochastic-dissipative classical dynamics of a charged particle 
(charge = $ -e, e>0$) and mass $m$ in the xy-plane in the presence of a uniform 
magnetic field of magnitude $B$ directed along the positive z-axis. 
Let the particle be confined harmonically in the xy-plane. 
(In this simple model, the motion along the z-axis factors out).  
Harder confinement (reflecting wall) will be introduced later.  
The equation of motion can now be written down straightforwardly in the polar 
co-ordinate system as
\begin{subequations}
\begin{equation}
m\ddot{r} = mr {\dot{\theta}}^2 - kr - \Gamma \dot{r} - {\frac{eB}{c}} r \dot{\theta} +  F_r(t,\Delta t)  
\end{equation}
\begin{equation}
mr\ddot{\theta} = -2m\dot{r} \dot{\theta} - \Gamma r \dot{\theta} + {\frac{eB}{c}} \dot{r} + F_{\theta}(t,\Delta t), 
\end{equation}
\end{subequations}
where, $F_r(t,\Delta t)$ and $F_{\theta}(t,\Delta t)$ are 
the Kubo-Anderson-type noise terms. The K-A noise, as depicted 
schematically in Fig. 1, 
is best described as a sequence of random rectangular pulses of equal
pulse-width $\Delta t$. The pulse heights are, however, identically, 
independently distributed gaussian random variates, of mean zero 
and finite variance.   

It is to be emphasized that $\Delta t$ here is 
not necessarily a small quantity in any sense
 $-$ it is to be regarded as a physical input parameter 
that makes the noise tunably  non-markovian.  The numerical 
simulation, however, must cover these not necessarily small 
time intervals $\Delta t$ with much finer sub-divisions $\delta t << \Delta t$ 
so as to ensure high numerical accuracy in treating the systematic parts of 
the driving forces present in Eqs. (1a) and (1b).  
More specifically, in our numerical 
simulation the number of finer time-step subdivisions has 
been taken to be $>500$ (Figures 1-3). 
Moreover, the  time of integration has been kept 
long ($~5\times 10^8$ sub-steps) in all 
these cases.

It is convenient to define here the parameters 
$\Omega_0$ $\equiv$ confining harmonic 
frequency = $\sqrt{k/m}$, $\Omega_c$ = the cyclotron frequency $eB/mc$, 
and $\gamma = \Gamma/m$, the Stokes friction. 
Also, we write that $F_r(t,\Delta t) = m\gamma^2\sigma f_r(t, \Delta t)$, 
and similarly for $F_\theta (t, \Delta t)$. 
Note that $\sigma$ has the dimension of length.  
Further, we introduce the dimensionless quantities 
$\omega_0 = \Omega_0/\gamma$,
$\omega_c = \Omega_c/\gamma$,
$\tau = \gamma t$,
and $R = r/\sigma$ the dimensionless radial coordinate.  
With these re-definitions, the equation 
(1a) and (1b) can now be re-written in the fully dimensionless form as
\begin{subequations}
\begin{equation}
\ddot{R} = R {\dot{\theta}}^2 - {\omega_0}^2 R - \dot{R} -  \omega_c R \dot{\theta} + f_r(\tau,\Delta \tau)   
\end{equation}
\begin{equation}
R\ddot{\theta} = -2\dot{R} \dot{\theta} - R \dot{\theta} + \omega_c \dot{R} + f_{\theta}(\tau,\Delta \tau),
\end{equation}
\end{subequations}
where $F_r(t,\Delta t) \equiv f_r(\tau, \Delta \tau)$, and 
$F_\theta(t,\Delta t) \equiv f_{\theta}(\tau, \Delta \tau)$.
Here the overhead dot denotes derivative with respect 
to the dimensionless time $\tau$.
Also, $f_r(\tau,\Delta \tau)$ and $f_{\theta}(\tau,\Delta \tau)$ are  
uncorrelated gaussian variates with mean zero and variance unity. 

The steady-state orbital magnetic moment $M$ can now be written as
\begin{equation}
M = \langle \langle -\frac{e}{2c} ({\bf r} \times \dot{\bf r})\rangle \rangle = - \left(\frac{e\gamma{\sigma}^2}{2c}\right) \langle \langle R^2\dot{\theta} \rangle \rangle
\end{equation}
giving the dimensionless orbital magnetic moment
\begin{equation}
\mu = \frac{M}{\left(\frac{e\gamma{\sigma}^2}{2c}\right)} = -\langle \langle R^2\dot{\theta} \rangle \rangle
\end{equation}
Here the double angle bracket denotes averaging over an ensemble of 
realizations of the random noise as well as over time in the long-time limit.  
We identify $\mu$ with the steady-state value of the dimensionless 
orbital magnetic moment. 

In figure 2 we have plotted the dimensionless magnetic magnetic 
moment $\mu$ against the pulse-width $\Delta \tau$, 
for some chosen values of the magnetic field, measured in terms 
of the dimensionless cyclotron frequency $\omega_c$.
The parameters $\omega_0$ and $\sigma$ have been set equal to unity. 
\begin{figure}[!t] 
\includegraphics[angle=-90,width=\columnwidth]{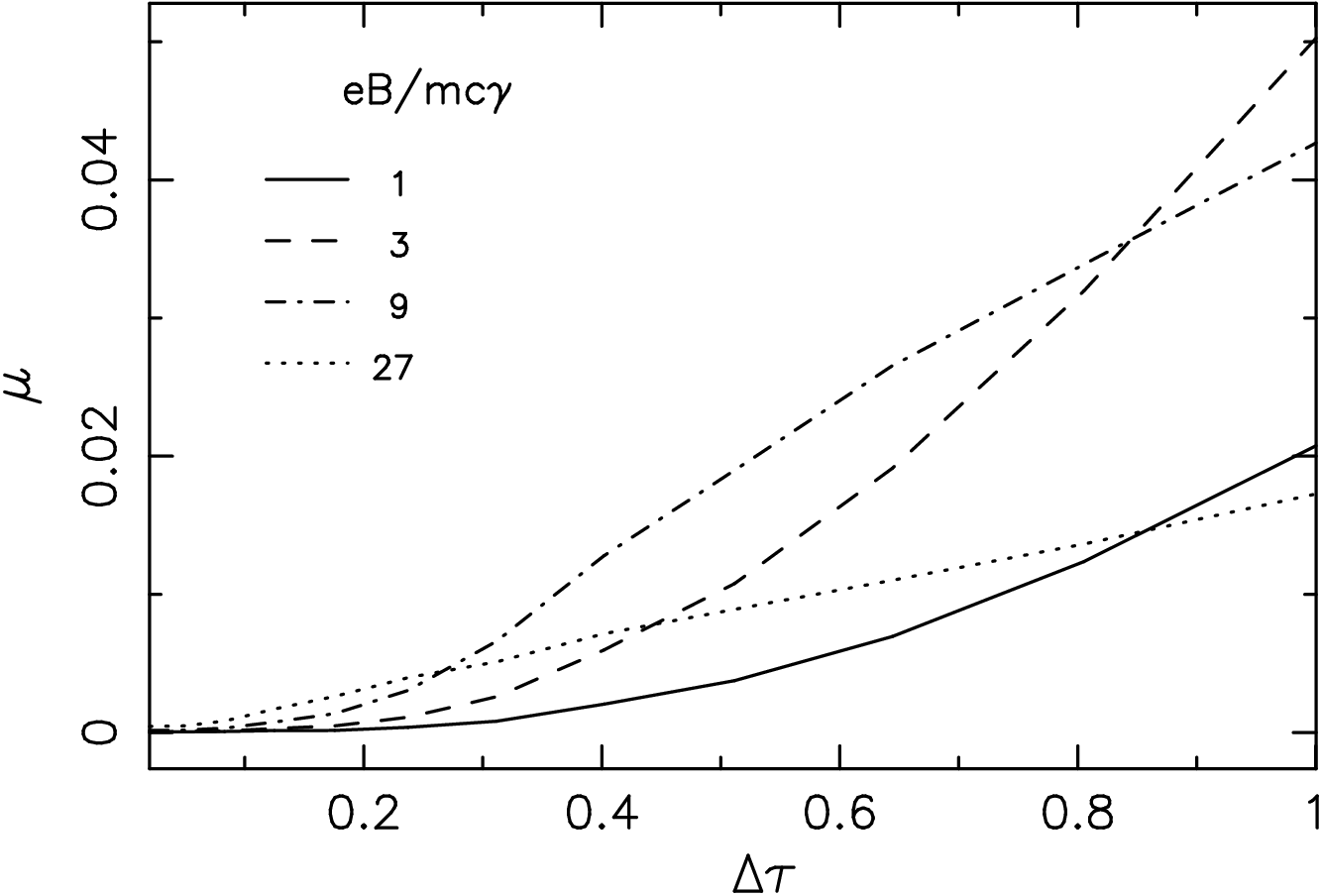}
\caption{Plot of dimensionless steady-state orbital magnetic moment 
$\mu$ against the pulse-width $\Delta \tau$ for some chosen values of 
the magnetic field (measured in terms of $\omega_c$).  
This is for the case of harmonic (soft) confinement. 
For details, see the text.}
\end{figure}

In figure 3, we have repeated, for the sake of comparison, the plot in
figure 2, but now for the case of a hard confinement. 
The latter is realized by modifying the soft, harmonic restoring 
force term $-kr$ in the radial equation (1a)
 to $-kr(r/a)^n$, and choosing a large integer exponent $n$.  The
length parameter $a$ here denotes the radial position 
of the effectively reflecting wall. With this, the term
$-{\omega_0}^2 R$ in
equation (2a)
becomes  $-{\omega_0}^2 R (R/A)^n$, with $A=a/\sigma$.
In figure 3 we have chosen n=20, and $A=1$. 
(Note that the limit $n=0$ corresponds
to the soft (harmonic) confining potential). 
\begin{figure}[!t] 
\includegraphics[angle=-90,width=\columnwidth]{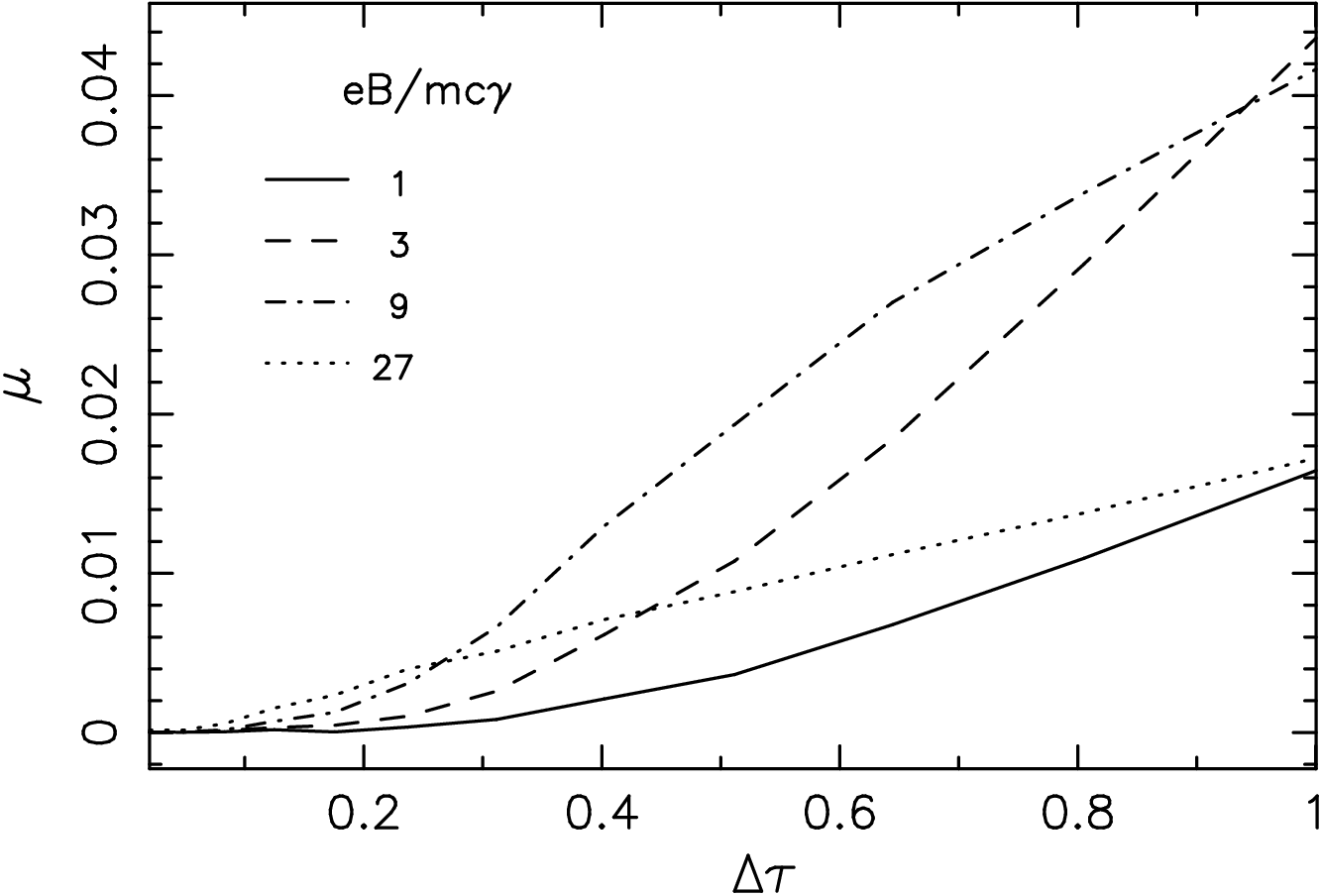}
\caption{Plot of dimensionless steady-state orbital magnetic moment 
$\mu$ against the pulse-width $\Delta \tau$ for some chosen values of 
the magnetic field (measured in terms of $\omega_c=eB/mc\gamma$).
This is for the case of hard confinement of nearly reflecting boundary 
as described in the text.}
\end{figure}

Finally, for completeness, in Fig. 4 we have repeated the plot 
(of $\mu$ against $\Delta \tau$) for the case of no confining potential, 
i.e., for $\omega_0 = 0$, with the other parameters remaining the same. 
\begin{figure}[!t] 
\includegraphics[angle=-90,width=\columnwidth]{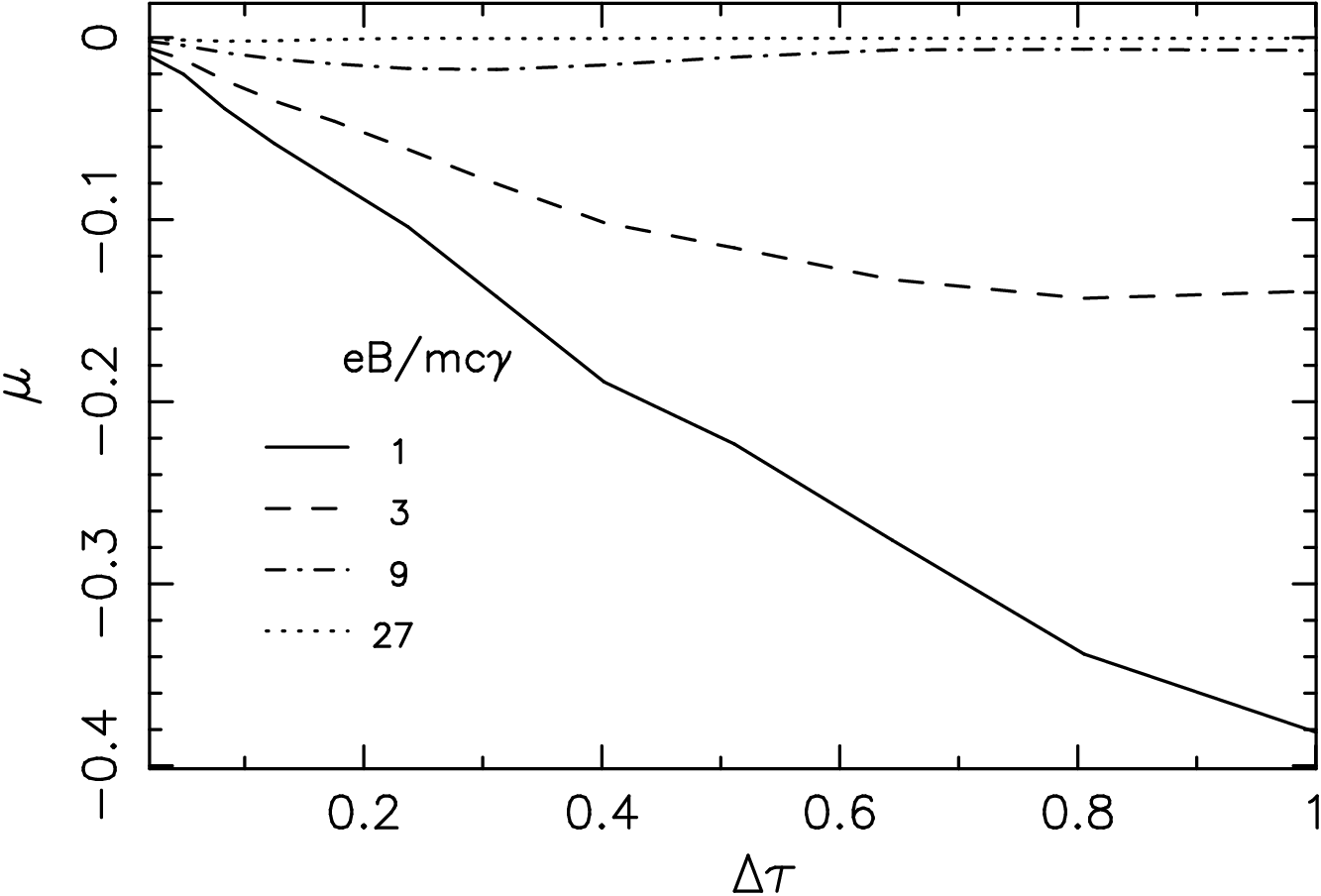}
\caption{ Plot of dimensionless steady-state orbital magnetic moment 
$\mu$ against the pulse-width $\Delta \tau$ for the same chosen values of 
parameters as in Figs. 2 and 3, except now for $\omega_0=0$ (no confinement).}
\label{Fig.4}
\end{figure}

\section{DISCUSSION}
Through our numerical simulation of a classical  model for the
stochastic-dissipative dynamics of a charged particle, moving  in a
magnetic field and driven by a  non-markovian  noise, we have
demonstrated the appearance  of an induced  orbital magnetic moment in
the  steady state. Most significantly, the orbital moment turns out to be
{\it paramagnetic}! The appearance of  a  non-zero classical orbital
magnetic moment in the non-equilibrium steady state implies deviation
of the dynamical system from the second fluctuation-dissipation
(II-FD) relation. The latter would have  otherwise  enforced  a
detailed-balance (meaning no cycles), and hence  no induced orbital  magnetic
moment. The moment, of course, vanishes in the limit of
delta-correlated white noise 
($i.e.,$ $\Delta\tau\rightarrow 0, \sigma^2\rightarrow \infty$, 
with the product $\Delta\tau \sigma^2$ finite) that renders 
the system II-FD theorem  compliant.  The essential point to be 
emphasized   here is the non-markovian
nature of the driving noise used in the stochastic  model. This point
is consistent with the proven result of Prost {\em et al.} (2009)
that for a
markovian dynamics, the  non-equilibrium steady state can always  be
transformed into an effectively equilibrium state.
It may be noted that our  results  hold  for a soft (harmonic)
potential confinement as well as for a hard (reflecting wall type)
potential  confinement.  Also, the magnitude of the orbital 
paramagnetic moment is non-monotonic in the externally applied 
magnetic-field  strength $-$ initially 
increasing with increasing magnetic field strength and then decreasing
in the high-field regime. Further, the magnetic moment scales as 
the variance of the random pulse-height
for a given pulse width in the case of harmonic confinement 
(see Eq. 3). In the case of the 
hard confinement, however, there is no simple scaling, 
but the qualitative behaviour remains the same. 

For the sake of completeness, we also carried out a similar simulation 
for the case of no potential-confinement,  $i.e.,$ with $\omega_0=0$. 
Interestingly, but not surprisingly though, we obtained as in 
figure 4 a steady-state orbital magnetic moment which has the 
opposite sign $-$ it is diamagnetic! This is, however, consistent 
with the physical picture of Bohr 
(Bohr 1911; van Leeuwen 1921; van Vleck 1932; Peierls 1979; Ma 1985):
without the confinement, 
there are naturally no orbits skipping the boundary, i.e., no edge current, 
which would have sub-tended a paramagnetic moment leading to the cancellation.  
We are thus left only with the Maxwell cycles well within the interior 
that constitute the amperean current loops giving the diamagnetic moment 
(the Lenz's Law). This avoided cancellation was seen also in the analytical 
solution for the simple case of a markovian (delta-correlated) 
noise (Jayannavar \& Kumar 1981).
This now turns out to be true even for the non-markovian 
case as shown in our simulation.  The overall picture is one in which 
the particle, initially at the origin, say, diffuses outwards on the 
average, but the orbital diamagnetic  moment reaches its steady-state 
value on a relatively short time scale which is determined by the 
parameters appearing in equations (1a,b). The outward diffusion 
simply spreads out the total orbital magnetic moment 
($\propto {\bf r}\times{\bf \dot{r}}$) over the ever increasing 
area covered, but without changing its time-averaged value $-$ the 
steady-state diamagnetic moment. 

\section{CONCLUDING REMARKS}
The paramagnetic sign of the induced orbital magnetic moment can have
interesting  consequences of considerable physical significance. After all,
the paramagnetic sign of the  moment inherently  signifies a positive
feed-back effect when we consider not just one but a system of many charged
particles. Here, the mean self-field can in principle lead to 
a spontaneous macroscopic orbital magnetic moment.

As for a possible experimental realization of such a confined system, 
we begin by noting that what is really essential for obtaining the 
classical orbital paramagnetism in a non-equilibrium steady state 
is the non-markovian nature of the stochastic forcing. 
Thus, a micron-sized sample of a semi-metal (such
as bismuth) trapped in an optical tweezer, and irradiated with  
random  laser pulses in the presence of an external magnetic field  
would constitute a possible candidate system. The
laser impulses should impart high enough kinetic energy (high nominal
temperature) so as to create a non-degenerate (classical) 
gas of charged particles (electrons and holes). 
Then, the high temperature washes away the quantum
signature, i.e., the discrete quantum level-spacings in
the micron-sized sample, leaving behind a classical charged particle system. 

Recall that the orbital magnetic moment does
not depend on the sign of the charge on the particle, e.g., be it an
electron or a hole. It is then reasonable to expect the
total induced orbital magnetic moment to scale up with the number of charged
particles in the confined system.

We would like to summarize now the main findings of our work 
as reported above. In the process, we hope to clarify some of the 
key points made therein. The classical (and indeed the classic) 
Bohr-van Leeuwen (BvL) theorem implies a complete absence of 
the equilibrium orbital magnetic moment inasmuch as the partition 
function then turns out to be independent of the magnetic 
{\it vector potential}. This global result has received an exact 
microscopic treatment based on the classical Langevin equation 
for the stochastic thermal motion of a charged (spinless) particle, 
the electron, in a static magnetic field, in the presence of friction 
(dissipation) and the random forcing (fluctuation) that satisfies 
the celebrated fluctuation-dissipation (F-D) theorem in thermal 
equilibrium. It turned out, however, that a deviation from the 
equilibrium F-D theorem did induce a {\it non-zero orbital magnetic 
moment in the system} (Kumar 2012). Moreover, it could be shown that the sign 
of the orbital magnetic moment (paramagnetic or diamagnetic) 
depends on the relative strengths of fluctuation and the 
dissipation effects that could be readily parametrized rather 
simply. Finally, these stochastic treatments based on the 
underlying Markovian noise is replaced here by the non-Markovian 
Kubo-Anderson process, providing finite correlation time-scale $\Delta \tau$
for noise driving without any memory from earlier states, 
and solved for now numerically. 

The present results, 
owing to the details of system and conditions
considered here being different,
do not render ready comparison with the earlier findings
across the different parameters probed in Kumar (2012). 
A limited comparison, however, is possible in the
dependence of the steady-state orbital magnetic moment
on the magnitude of the magnetic field strength.
More specifically, as can be viewed from the set of curves
in Figures 2 \& 3,
the orbital magnetic moment at any given $\Delta \tau$
first increases with magnetic field, and then reduces with
further increase in the magnetic field.  
This behaviour is qualitatively consistent with the trend apparent 
in the Figure 1 of Kumar (2012) for the paramagnetic signature
(i.e., when $\eta$=0.5). 
The Kubo-Anderson process 
(where the transition rates are independent of original state 
the transition came from) is admittedly more realistic (physical) 
than the usual gaussian processes.
  
It is our conjecture that this non-zero classical orbital 
Paramagnetism of the non-equilibrium steady-state, 
driven by the Kubo-Anderson stochastic noise, 
should hold true in general across a wider range of situations. 
If indeed it turns out to be the case, it would be a surprise of 
theoretical physics akin to the Bohr--van Leeuwen (BvL) 
theorem for the equilibrium quantum mechanics! 
This mechanism may even 
generate spontaneously a macroscopic magnetic meanfield $-$ such 
as the {\em seed} field of interest in the astrophysical context. 

One of the issues central to the Compact Stars (related astrophysics) 
is the origin of magnetic effects (magnetic fields) associated with
the stellar objects\footnote{Magnetic fields have indeed threaded
Srini's works on compact stars, particularly that elucidating
the fine details of evolution of both and 
the remarkable interplay.}, in particular their electron-gas 
effects due to the {\it motion} of the charged 
particles (electrons). This {\it orbital} (not the spin) effect
is precisely what has been addressed in our work!
The idea underlying our work is that a deviation from the F-D 
(Fluctuation-Dissipation (theorem))  naturally leads to 
a magnetic effect (susceptibility) that can indeed be
paramagnetic in a {\it confined} system of charged particles (electrons),
without involving their spin. It ultimately  depends on which 
of the two --the Fluctuations or the Disspation --dominates!
In our model, this is controlled by the stochastic parameters,
such as the finite correlation time scale of the random forcing
(in comparison with the cyclotron period).

\end{document}